\documentstyle[12pt,emlines2]{article}
\newcommand{\bm}[1]{\mbox{\boldmath$#1$}}

\newcommand{\na}{\mbox{\boldmath$\nabla$}}
\newcommand{\vv}{\mbox{\boldmath$v$}}
\newcommand{\EE}{\mbox{\boldmath$E$}}
\newcommand{\FF}{\mbox{\boldmath$F$}}

\newcommand{\rr}{\mbox{\boldmath$r$}}

\newcommand{\uu}{\mbox{\boldmath$u$}}
\newcommand{\HH}{\mbox{\boldmath$H$}}
\begin{document}

\begin{center}
{\bf Uncertainty Relations (UR)
Have Nothing to do with Quantum Mechanics (QM)}
\bigskip

Vladimir K. Ignatovich
\bigskip

Joint Institute for Nuclear Research
Laboratory of Neutron Physics
Dubna, Russia
\bigskip

\begin{abstract}
Uncertainty relations are shown to have nothing specific for
quantum mechanics, being the general property valid for arbitrary
function. A wave function of a particle having precisely defined
position and momentum in QM simultaneously is demonstrated.
Interference on two slits in a screen is shown to exist in
classical mechanics. A nonlinear classical system of
equations replacing QM Schr\"odinger equation is suggested.
This approach is shown to have nothing in common with Bohmian mechanics.
\end{abstract}
\end{center}

\section{Introduction}

The title of the paper is not truth, because uncertainty relations are valid in QM,
however we chose this title in a protest against the widely spread belief that UR are the cornerstone
of QM. The real cornerstone of QM is the Schr\"odinger equation, which was a great guess, like Maxwellian ones.
In the next section we remind to readers how UR are proven for an arbitrary function.
It follows from this proof that UR have nothing specific for QM.
In the third section we show that the such notions as position and momentum are a matter of definition for
an extended object like a wave function, and demonstrate that nonsingular de Broglie wave packet describes a
particle, which simultaneously has precisely defined momentum and position. In the forth section we show
that interference is not an exclusive property of a wave mechanics. It takes place also in classical mechanics.
In fifth section we discuss whether QM equation can be replaced with classical equations. We suppose that
it is possible to define a system of equations for trajectory and field of the particle, propose for mathematicians
to solve an electrodynamical problem for an electron moving through a slit in a conducting screen, and show
that such a system of equations is not contained in so called ``Bohmian mechanics''.
And in conclusion we repeat our main points.

\section{What are UR}

UR is a mathematical theorem which relates ranges of a function and its Fourier image.
This theorem is valid in all branches of physics and mathematics dealing with extended objects described with
functions. Let us remind this well known theorem.

Take an arbitrary function $f(x)$ of finite range, and its
Fourier image
\begin{equation}\label{1}
F(p)=\int\limits_{-\infty}^{+\infty}f(x)\exp(ipx)dx,
\end{equation}
and define
\begin{equation}\label{2}
\int\limits_{-\infty}^{+\infty}|f(x)|^2dx=\int\limits_{-\infty}^{+\infty}|F(p)|^2dp\equiv N<\infty,
\end{equation}
\begin{equation}\label{3}
x_0=\frac1N\int\limits_{}^{}x|f(x)|^2dx,\qquad p_0=\frac1N\int\limits_{}^{}p|F(p)|^2dp,
\end{equation}
With this function we can write the nonnegative integral
\begin{equation}\label{4}
\frac1N\int\limits_{}^{}|(\alpha(x-x_0)+d/dx-ip_0)f(x)|^2=
\alpha^2A+\alpha B+C
\end{equation}
for arbitrary $\alpha$, where
\begin{equation}\label{5}
A=\frac1N\int\limits_{}^{}(x-x_0)^2|f(x)|^2dx=\frac1N\int\limits_{}^{}(x^2-x_0^2)|f(x)|^2dx\equiv<(\Delta x)^2>
\end{equation}
\begin{equation}\label{6}
B=\frac1N\int\limits_{}^{}x\frac{d}{dx}|f(x)|^2dx=\int\limits_{}^{}\frac{d}{dx}
(x|f(x)|^2)-\int\limits_{}^{}|f(x)|^2=-\frac1N\int\limits_{}^{}|f(x)|^2=-1
\end{equation}
\begin{equation}\label{7}
C=\frac1N\int\limits_{}^{}(p-p_0)^2|F(p)|^2dp=\frac1N\int\limits_{}^{}(p^2-p_0^2)|F(p)|^2dp\equiv<(\Delta p)^2>.
\end{equation}
Since eq. (\ref{4}) is nonnegative for all $\alpha$, we have
$$\alpha^2<(\Delta x)^2>-\alpha +<(\Delta p)^2>\ge0$$
which is possible only for
\begin{equation}\label{1b}
<(\Delta p)^2><(\Delta x)^2>\ge\frac{1}{4},
\end{equation}
which is just the uncertainty relation used in QM, however it is satisfied for arbitrary function $f(x)$,
and therefore is not related specifically to QM. Thus it cannot be a corner stone of QM.
The uncertainty relation takes place in all branches of physics. For example, in
classical field theory,
thermodynamics,
hydrodynamics,
plasma.
It is valid even in classical mechanics, because for functions $x(t)$
we have UR $(\Delta \omega)^2(\Delta t)^2\ge 1/4$.

{\bf UR contain nothing, specific to QM.
QM is only a particular case, which is very
alike to classical field theory.}

\section{Position and momentum can be defined absolutely precisely
simultaneously}

Since the wave function in QM defines a particle and it is an extended object, the question
arises: what is a position of the extended object?

The answer to this question is: position of the extended object is the matter of definition.

In classical electrodynamics position of the electron is the singularity
of its field.

In classical mechanics position of, say, a ball is the matter of definition. You may choose its
center or a point, where you touch it.

For a free particle of mass $m$ in QM we can use the nonsingular de Broglie's wave-packet~\cite{bro,ig1,ig2}:
\begin{equation}\label{8}
\psi=j_0(s|\bm{r-v}t|)\exp(i\bm{vr}-i\omega t),
\end{equation}
in which $j_0(x)$ is the spherical Bessel function, $s$ is a parameter determining the width of the function, and
\begin{equation}\label{9}
\omega=(v^2+s^2)/2.
\end{equation}
Here we use unities $\hbar=m=1$, so velocity $v$ of the particle is the same as its wave-vector $k$.
The function (\ref{9}) is a solution of the Schr\"odinger equation:
$$(i\partial_t+\Delta/2)\psi=0.$$
We can define position of it as the position of maximum of $|\psi|^2$ and as a momentum
it's velocity $v$. They are defined absolutely precisely simultaneously in QM.

\section{Interference in classical mechanics}

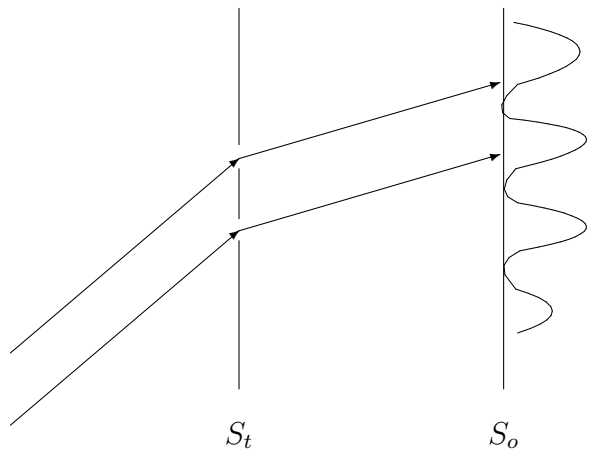
\begin{figure}[bht]
\begin{center}
\special{em:linewidth 0.4pt}
\unitlength 0.80mm
\linethickness{0.4pt}
\begin{picture}(109.00,72.67)
\emline{39.00}{72.67}{1}{39.00}{50.00}{2}
\emline{39.00}{46.00}{3}{39.00}{37.67}{4}
\emline{39.00}{34.00}{5}{39.00}{9.34}{6}
\emline{83.00}{72.67}{7}{83.00}{9.34}{8}
\put(39.00,35.67){\vector(4,3){0.2}}
\emline{1.00}{3.34}{9}{39.00}{35.67}{10}
\put(82.33,48.34){\vector(3,1){0.2}}
\emline{39.00}{35.67}{11}{82.33}{48.34}{12}
\put(39.00,47.67){\vector(4,3){0.2}}
\emline{1.00}{15.34}{13}{39.00}{47.67}{14}
\put(82.33,60.34){\vector(3,1){0.2}}
\emline{39.00}{47.67}{15}{82.33}{60.34}{16}
\emline{84.67}{70.34}{17}{86.99}{69.84}{18}
\emline{86.99}{69.84}{19}{89.04}{69.33}{20}
\emline{89.04}{69.33}{21}{90.81}{68.81}{22}
\emline{90.81}{68.81}{23}{92.31}{68.28}{24}
\emline{92.31}{68.28}{25}{93.53}{67.74}{26}
\emline{93.53}{67.74}{27}{94.48}{67.18}{28}
\emline{94.48}{67.18}{29}{95.15}{66.62}{30}
\emline{95.15}{66.62}{31}{95.55}{66.05}{32}
\emline{95.55}{66.05}{33}{95.67}{65.46}{34}
\emline{95.67}{65.46}{35}{95.51}{64.87}{36}
\emline{95.51}{64.87}{37}{95.08}{64.26}{38}
\emline{95.08}{64.26}{39}{94.37}{63.65}{40}
\emline{94.37}{63.65}{41}{93.39}{63.02}{42}
\emline{93.39}{63.02}{43}{92.14}{62.38}{44}
\emline{92.14}{62.38}{45}{90.60}{61.73}{46}
\emline{90.60}{61.73}{47}{88.80}{61.07}{48}
\emline{88.80}{61.07}{49}{85.33}{60.00}{50}
\emline{85.33}{60.00}{51}{83.50}{58.14}{52}
\emline{83.50}{58.14}{53}{82.67}{56.58}{54}
\emline{82.67}{56.58}{55}{82.83}{55.31}{56}
\emline{82.83}{55.31}{57}{84.00}{54.34}{58}
\emline{84.00}{54.34}{59}{86.39}{54.06}{60}
\emline{86.39}{54.06}{61}{88.53}{53.77}{62}
\emline{88.53}{53.77}{63}{90.42}{53.46}{64}
\emline{90.42}{53.46}{65}{92.07}{53.14}{66}
\emline{92.07}{53.14}{67}{93.47}{52.81}{68}
\emline{93.47}{52.81}{69}{94.62}{52.45}{70}
\emline{94.62}{52.45}{71}{95.52}{52.09}{72}
\emline{95.52}{52.09}{73}{96.18}{51.71}{74}
\emline{96.18}{51.71}{75}{96.59}{51.31}{76}
\emline{96.59}{51.31}{77}{96.75}{50.90}{78}
\emline{96.75}{50.90}{79}{96.67}{50.47}{80}
\emline{96.67}{50.47}{81}{96.33}{50.03}{82}
\emline{96.33}{50.03}{83}{95.75}{49.57}{84}
\emline{95.75}{49.57}{85}{94.92}{49.10}{86}
\emline{94.92}{49.10}{87}{93.85}{48.61}{88}
\emline{93.85}{48.61}{89}{92.53}{48.10}{90}
\emline{92.53}{48.10}{91}{90.96}{47.59}{92}
\emline{90.96}{47.59}{93}{89.14}{47.05}{94}
\emline{89.14}{47.05}{95}{87.07}{46.50}{96}
\emline{87.07}{46.50}{97}{85.00}{46.00}{98}
\emline{85.00}{46.00}{99}{83.52}{44.14}{100}
\emline{83.52}{44.14}{101}{83.08}{42.58}{102}
\emline{83.08}{42.58}{103}{83.69}{41.31}{104}
\emline{83.69}{41.31}{105}{85.33}{40.34}{106}
\emline{85.33}{40.34}{107}{87.66}{39.90}{108}
\emline{87.66}{39.90}{109}{89.73}{39.47}{110}
\emline{89.73}{39.47}{111}{91.53}{39.03}{112}
\emline{91.53}{39.03}{113}{93.06}{38.60}{114}
\emline{93.06}{38.60}{115}{94.33}{38.16}{116}
\emline{94.33}{38.16}{117}{95.33}{37.73}{118}
\emline{95.33}{37.73}{119}{96.07}{37.29}{120}
\emline{96.07}{37.29}{121}{96.54}{36.86}{122}
\emline{96.54}{36.86}{123}{96.74}{36.42}{124}
\emline{96.74}{36.42}{125}{96.68}{35.99}{126}
\emline{96.68}{35.99}{127}{96.35}{35.55}{128}
\emline{96.35}{35.55}{129}{95.76}{35.12}{130}
\emline{95.76}{35.12}{131}{94.90}{34.68}{132}
\emline{94.90}{34.68}{133}{93.78}{34.25}{134}
\emline{93.78}{34.25}{135}{92.39}{33.81}{136}
\emline{92.39}{33.81}{137}{90.73}{33.38}{138}
\emline{90.73}{33.38}{139}{88.81}{32.95}{140}
\emline{88.81}{32.95}{141}{85.67}{32.34}{142}
\emline{85.67}{32.34}{143}{83.88}{31.19}{144}
\emline{83.88}{31.19}{145}{83.05}{29.87}{146}
\emline{83.05}{29.87}{147}{83.19}{28.37}{148}
\emline{83.19}{28.37}{149}{85.00}{26.00}{150}
\emline{85.00}{26.00}{151}{87.16}{25.26}{152}
\emline{87.16}{25.26}{153}{88.85}{24.53}{154}
\emline{88.85}{24.53}{155}{90.07}{23.80}{156}
\emline{90.07}{23.80}{157}{90.81}{23.07}{158}
\emline{90.81}{23.07}{159}{91.08}{22.33}{160}
\emline{91.08}{22.33}{161}{90.88}{21.60}{162}
\emline{90.88}{21.60}{163}{90.20}{20.87}{164}
\emline{90.20}{20.87}{165}{89.05}{20.13}{166}
\emline{89.05}{20.13}{167}{87.43}{19.40}{168}
\emline{87.43}{19.40}{169}{85.33}{18.67}{170}
\put(39.00,1.67){\makebox(0,0)[cc]{$S_t$}}
\put(83.00,1.67){\makebox(0,0)[cc]{$S_o$}}
\end{picture}
\caption{\label{f1}According to standard QM wave function of a particle transmitted
through both slits in target screen $S_t$ interferes after $S_t$ and gives
a diffraction pattern on observation screen $S_o$.}
\end{center}
\end{figure}

Let's consider an experiment on interference on two slits in a screen, shown in fig. \ref{f1}.
It is usually stated that particle goes through both slits in the screen,
and transmitted parts of the particle wave function interfere on the screen of observation,
which is manifested by the interference pattern.
However the interference can be explained purely classically with
particle going through only one exactly specified slit.

\begin{figure}[thb]
\begin{center}
\special{em:linewidth 0.4pt}
\unitlength 0.80mm
\linethickness{0.4pt}
\begin{picture}(89.00,75.00)
\emline{38.33}{70.33}{1}{38.33}{47.66}{2}
\emline{38.33}{43.66}{3}{38.33}{35.33}{4}
\emline{38.33}{31.66}{5}{38.33}{7.00}{6}
\put(38.33,45.33){\vector(4,3){0.2}}
\emline{0.33}{13.00}{7}{38.33}{45.33}{8}
\put(81.66,58.00){\vector(3,1){0.2}}
\emline{38.33}{45.33}{9}{81.66}{58.00}{10}
\emline{83.33}{75.00}{11}{83.33}{6.66}{12}
\put(89.00,58.00){\makebox(0,0)[lc]{2-nd slit closed}}
\put(89.00,45.66){\makebox(0,0)[lc]{2-nd slit open}}
\put(35.66,24.66){\rule{1.33\unitlength}{7.00\unitlength}}
\put(36.33,23.66){\vector(0,1){0.2}}
\emline{36.33}{15.66}{13}{36.33}{23.66}{14}
\put(81.67,45.33){\vector(1,0){0.2}}
\emline{38.33}{45.33}{15}{81.67}{45.33}{16}
\put(38.33,1.67){\makebox(0,0)[cc]{$S_t$}}
\put(83.33,1.33){\makebox(0,0)[cc]{$S_o$}}
\put(29.33,26.33){\makebox(0,0)[cc]{sh}}
\put(18.67,32.67){\makebox(0,0)[cc]{e}}
\end{picture}
\caption{\label{f2}An experiment with classical electron going through the upper slit
in the screen $S_t$. Because of interaction of the electron field with the
$S_t$ its trajectory after the screen depends on whether the other slit
is opened or not. It is an illustration of interference of two slits in
classical physics.}
\end{center}
\end{figure}
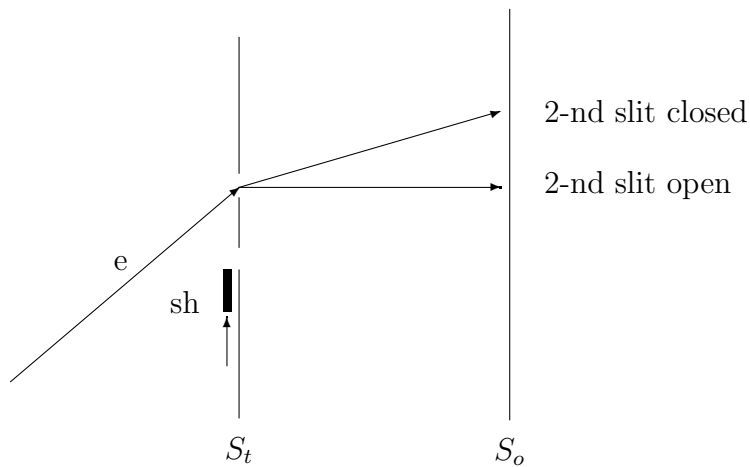

Let us consider the same experiment with a classical electron, moving through
one specified slit in the target screen, as is shown in fig. \ref{f2}.
Because of interaction of the electron field with the screen, the electron
trajectory changes after the screen. Interaction of the electron field
with the screen $S_t$ depends on the screen structure. In particularly, it
is different when there is one or two slits. It means
that the direction of propagation of the electron after $S_t$ depends on
whether the second slit is opened or closed. Thus the second slit interfere
with electron motion, even if the electron goes precisely through the chosen
upper slit.

Our considerations permit to predict the change of direction of the electron
after the screen $S_t$, if we perform an experiment shown in fig. \ref{f2},
where the second slit can be closed with the shutter sh. With such simple
considerations we cannot predict the diffraction pattern on the screen $S_o$,
shown in fig. \ref{f1}, because in classical physics there are no such a parameter as
wave-length, however wavelength can enter, if we take into account
relativistic retardation of the interaction of electron with its own field
reflected from the screen $S_t$ or introduce a quantum of action.
Indeed, we can suppose that the shift of the incident electron
along distance $l$ can affect the total field of the electron in presence of the screen $S_t$,
and consecutively electron motion only if $pl=h$. Just at this point the quantization can enter
into the classical behavior, and give such a parameter as the wave-length.

\section{Nonlinear classical system of equations instead of QM}

All the usual equations in mathematical physics can be sorted into two groups:
\begin{enumerate}
\item
{\bf  Field equations} of the type
\begin{equation}\label{3a}
\hat L\psi(\rr)=j(\rr),
\end{equation}
where $\hat L$ is an operator, which can be linear or nonlinear in field $\psi(\rr)$,
and $j(\rr)$ is a source, which can depend on some particle trajectory
$\rr(t)$, and this trajectory is supposed to be fixed. As an example we can
mention Maxwell equations with given currents, and with
determined boundary conditions.
\item {\bf Trajectory equations} of the type
\begin{equation}\label{4a}
\frac{d^2\rr}{dt^2}=\FF(\rr(t),t),
\end{equation}
where the field of force $\FF(\rr,t)$ is fixed.
\end{enumerate}

However, above, we had another type of the problem. It differs from (\ref{3a}) and (\ref{4a}).
In this problem one has a trajectory equation
\begin{equation}
\frac{d^2\rr_p}{dt^2}=\FF(\psi(\rr_p(t),t)),
\label{1a}
\end{equation}
with the force $\FF(\psi)$, which depends on unknown field $\psi$.
The field $\psi$ is a solution of the field equation
\begin{equation}
\hat L\psi(\rr,t)=j(\rr,\rr_p(t))
\label{2a}
\end{equation}
with the source, which depends on yet unknown solution of equation
(\ref{1a}).

Formally we can exclude $\psi=\hat L^{-1}j(\rr,\rr_p(t))$ from the
equation (\ref{1a}), however then we obtain highly nonlinear equation for trajectory:
\begin{equation}
\frac{d^2\rr_p}{dt^2}=\FF(\hat L^{-1}j(\rr_p(t),\rr_p(t'))).
\label{5a}
\end{equation}
Solution of (\ref{5a}) or of the system (\ref{1a},\ref{2a})
is the {\bf challenge for mathematicians}.

QM avoids solution of such a
nonlinear system, however reduction of nonlinear system to a
linear Schr\"odinger equation costs probabilities instead of
determinism.

However it would be very interesting to try to solve such a nonlinear system,
which can be easily formulated in classical electrodynamics.

\subsection{The problem of classical electrodynamics}

We have the Maxwell equation for 4-tensor $\FF_{\mu\nu}$:
\begin{equation}
\partial_\mu F_{\mu\nu}(\rr,t)=\frac{4\pi}{c}eu_\nu\delta(\rr-\rr(t)),
\quad \mu,\nu=0\div3,
\label{a}
\end{equation}
where $u_\nu$ is speed with components $u_0=c$, $\uu_k=\vv_k(t)$ for $k=1\div3$.
The functions $\rr(t)$ and $\vv(t)$ are not known and are to be determined
from the other equation --- the trajectory one:
$$m\frac{d\vv(t)}{dt}=e\EE(\rr,t)+\frac{e}{c}[\vv(t)\HH(\rr,t)],$$
where
$$\vv(t)=\frac{d\rr(t)}{dt},$$
and electric and magnetic fields are the components of the 4-tensor
$\FF_{\mu\nu}$
$$\quad \EE_k(\rr,t)=F_{0k}(\rr,t), \quad
\HH_k=\epsilon_{ijk}F_{ij}(\rr,t),$$
which are formed by field, reflected from the target screen, and the
reflection is determined by
boundary conditions for the field $F_{\mu\nu}$.
The screen can be accepted to be an infinitely thin
ideal conductor. Position of slits, their width and the distance between them
can be arbitrary.

For the beginning it is sufficient to solve even non relativistic,
pure Coulomb problem.
In the case, when there are no slits, solution in
nonrelativistic limit is trivial.

We want to remark that this nonlinear system has nothing to do with Bohmian mechanics.
No quantum potential is introduced, and no Schr\"odinger equation is presupposed.
In the next section we briefly review the Bohmian mechanics.

\subsection{Bohmian mechanics and hydrodynamical interpretation}

There are a lot of activity on interpretation of quantum mechanics
in terms of classical trajectories and quantum potential (see, for
example,~\cite{bohm} and references there in), which are known as
Bohmian mechanics or hydrodynamical interpretation. However it is
not a classical version, which replaces quantum mechanics, but
only an alternative way of solving of the Schr\"odinger equation. We
can find the full wave function $\psi$ by solving the
Schr\"odinger equation
\begin{equation}\label{bom}
i\hbar\frac{\partial}{\partial t}\psi(\rr,t)=\left[-\frac{\hbar^2}{2m}\Delta+V(\rr)\right]\psi,
\end{equation}
or represent it as $\psi(\rr,t)=R(\rr,t)\exp(iS(\rr)/\hbar)$, where
$R(\rr)=|\psi(\rr)$,
substitute into (\ref{bom}), and after separation of real and imaginary parts obtain
two other equations~\cite{bohm} for them
\begin{equation}\label{bom1}
\frac{\partial R^2}{\partial t}+\na\left(R^2\frac{\na S}{m}\right)=0,
\end{equation}
\begin{equation}\label{bom2}
\frac{\partial S}{\partial t}+\frac{(\na S)^2}{2m}+V-\frac{\hbar^2}{2m}\frac{\na^2R}{R}=0.
\end{equation}
Solution of these two equations is equivalent to solution of the single (\ref{bom})
equation. When you find $R$ and $S$, you can find such things as
\begin{equation}\label{bom3}
Q(\rr,t)=\frac{\hbar^2}{2m}\frac{\na^2R}{R},
\end{equation}
which you call ``quantum potential'', and
\begin{equation}\label{bom4}
\vv(\rr,t)=\frac{\na S(\rr,t)}{m}
\end{equation}
which you call speed. If you apply $\na$ to Eq. (\ref{bom2}) and use definition
(\ref{bom4}), you obtain the equation
\begin{equation}\label{bom5}
m\frac{\partial\vv(\rr,t)}{\partial t}+m(\vv\cdot\na)\vv=-\na(V(\rr)+Q(\rr,t)),
\end{equation}
which is equivalent to
\begin{equation}\label{bom6}
\frac{d\vv(\rr,t)}{dt}=-\na(V(\rr)+Q(\rr,t)).
\end{equation}
However $\vv$ is not equal to $\dot{\rr}(t)$, because it is a field, which depends on
both $\rr$ and $t$.

Now, if you have already solved  Eq. (\ref{bom}), you can consider (\ref{bom6}) as
the Newton equation and find a family of trajectories. However, in this case you
arrive at the problem of finding trajectories for given field (\ref{4a}).
It has nothing in common with the proposed classical nonlinear system of equations.

\section{Conclusion}

We think that the wave function $\psi$ in QM represents some kind of a
field, and the force of this field can be proportional to
$|\psi|^2$. Then it will explain why in QM probability for a
particle to be detected is proportional to $|\psi|^2$. If $\psi$
is a field, then the position and momentum of a particle which is the source of
this field can be naturally defined simultaneously, and UR do not
forbid it.

\section{A history of referee reports}
I think the science now is very alike to a religion, and because
of this there is a strong theocratic like censorship. I am sure
that publication of this paper will be almost impossible. I want
to start here the history of all the referee reports and my
replies to them. May be it will be useful for history of science.

\subsubsection{From Phys.Lett. A}
Dear Dr Ignatovich,

On 05-Mar-2004 you submitted a new paper number
PLA-ignatovi.AT.jinr.ru-20040305/1 to Physics Letters A entitled:

``Uncertainty Relations (UR) Have Nothing to do with Quantum
Mechanics (QM)''

We regret to inform you that your paper has not been accepted for
publication. Comments from the editor are:

I enclose a report on your paper. In view of the referee's remarks
I regret that your article is not suitable for publication in
Physics Letters A.

Thank you for submitting your work to our journal.

--

Report

1)The fact that mathematical relations having the form of the
uncertainty relations occur in many fields of physics does not
imply that the uncertainty relations have nothing to do with
quantum mechanics, as claimed in the title.

2)The remarks of sections 2, 3 and 5, where they are correct, are
trivial and well known.

3)The claim to give a classical model in section 4 is just
hand-waving. It is well known that one can retain the particle
trajectory in the two-slit experiment and there is a huge
literature on this.

4)A reference which touches on the points raised in this paper,
and deals with them with considerably more rigour, detail and
consistency, is P. Holland, The Quantum Theory of Motion
(Cambridge UP, 1993).

The paper should be rejected.

--

Yours sincerely,

J.P. Vigier Editor Physics Letters A For queries about the
E-submission website please contact: authorsupport@elsevier.com
For queries specific to this submission please contact: c/o
phys.letters@green.oxford.ac.uk

\paragraph{My comment} Though the report is absolutely negative and does not suppose
a discussion, we try to improve our paper.

1)In the first section we included a sentence, that UR are not a
corner stone for QM. The real corner stone is the Schr\"odinger
equation, which was genuinely guessed. We explained also, why did we choose such a title.

2)According to this remark my sections 2, 3 and 5 are trivial,
because there are no incorrect remarks. So, the referee agrees
that the position and momentum of a particle can be defined
simultaneously. In that case he must accept that UR do not forbid
it. Iam grateful to the referee for his support.

3)I agree that my model in section 4 is in some respect hand waving. However I
can precisely formulate a mathematical problem for solution. And I
included it in the section 5. As for trajectories in Bohmian
mechanics, I included a paragraph also.

4)Thanks to referee for the reference. I included another
one~\cite{bohm}, more recent, which is related to the problem and
to the point 3) of the referee report.


\begin{thebibliography}{99}
\bibitem{bro}
L. de Broglie, {\it Non-Linear Wave Mechanics: A
Causal Interpretation.} (Elsevier, Amsterdam, 1960).
\bibitem{ig1}V.K.Ignatovich,
{\it The Physics of Ultracold Neutrons (UCN). Oxford
Clarendon Press. 1990.}
\bibitem{ig2}
Ignatovich V.K., Utsuro M.
Tentative solution of UCN problem. PL A 255 (4-6) p. 195-202 1997.
\bibitem{bohm}
A.S.Sanz, F.Borondo, and S.Miret-Artes, Particle diffraction
studied using quantum trajectories. J.Phys.: Condens. Matter, 14,
p. 6109-6145, 2002.
\end{thebibliography}
\end{document}